\documentclass{ws-ijmpd}

\begin{document}

\def\nocropmarks{\vskip5pt\phantom{cropmarks}}

\let\trimmarks\nocropmarks      

\markboth{R. Ruffini, F. Fraschetti, L. Vitagliano, S.-S. Xue}{Observational signatures of an electromagnetic overcritical gravitational collapse}

\catchline{}{}{}

\title{Observational signatures of an electromagnetic overcritical gravitational collapse}

\author{\footnotesize REMO RUFFINI}
\address{ICRA --- International Center for Relativistic Astrophysics and Dipartimento di Fisica,\\ Universit\`a di Roma ``La Sapienza'', Piazzale Aldo Moro 5, I-00185 Roma, Italy.}

\author{\footnotesize FEDERICO FRASCHETTI}
\address{ICRA --- International Center for Relativistic Astrophysics and Dipartimento di Fisica,\\Universit\`a di Trento, Via Sommarive 14, I-38050 Povo (Trento), Italy.}

\author{\footnotesize LUCA VITAGLIANO}
\address{ICRA --- International Center for Relativistic Astrophysics and Dipartimento di Fisica,\\ Universit\`a di Roma ``La Sapienza'', Piazzale Aldo Moro 5, I-00185 Roma, Italy.}

\author{\footnotesize SHE-SHENG XUE}
\address{ICRA --- International Center for Relativistic Astrophysics and Dipartimento di Fisica,\\ Universit\`a di Roma ``La Sapienza'', Piazzale Aldo Moro 5, I-00185 Roma, Italy.}

\maketitle

\begin{history}
\received{Day Month Year}
\revised{Day Month Year}
\end{history}

\begin{abstract}
We present theoretical predictions for the spectral, temporal and intensity signatures of the electromagnetic radiation emitted during the process of the gravitational collapse of a stellar core to a black hole, during which electromagnetic field strengths rise over the critical value for $e^+e^-$ pair creation. The last phases of this gravitational collapse are studied, leading to the formation of a black hole with a subcritical electromagnetic field, likely with zero charge, and an outgoing pulse of initially optically thick $e^+e^-$-photon plasma. Such a pulse reaches transparency at Lorentz gamma factors of $10^2$--$10^4$. We find a clear signature in the outgoing electromagnetic signal, drifting from a soft to a hard spectrum, on very precise time-scales and with a very specific intensity modulation. The relevance of these theoretical results for the understanding of short gamma-ray bursts is outlined.
\end{abstract}

\keywords{EMBH --- electron-positron plasma --- gravitational collapse --- gamma-ray bursts}

\section{Introduction}
The discovery in 1997 of the afterglows of Gamma-Ray Bursts (GRBs) \cite{C97} has evidenced the cosmological nature of these sources. By the analysis of the first and second BATSE catalogs\footnote{see http://cossc.gsfc.nasa.gov/batse/} Tavani in 1998 \cite{T98} confirmed the existence of two families of GRBs: the so-called ``long-bursts'' with a soft spectrum and duration $\Delta t > 2.5$sec and the ``short-bursts'' with harder spectrum and duration $\Delta t < 2.5$sec. In 2001 the theory was advanced \cite{lett2} that both short-bursts and long-bursts originate from the same underlying physical process due to the vacuum polarization of electromagnetic overcritical gravitational collapse leading to the creation of $e^+-e^-$ pairs at the expenses of the extractable energy of a black hole \cite{CR71}. The difference between the short-bursts and long-bursts in this theory is mainly due to the amount of baryonic matter encountered by the $e^+e^-$ pairs in their relativistic expansion. A support of such a theory was given by Schmidt \cite{S01} showing that short-bursts and long-bursts have the same isotropic-equivalent characteristic peak luminosity.

In recent work we have systematically developed the theoretical background of a process
of gravitational collapse of matter 
involving an electromagnetic field with field strength higher than the
critical value for $e^+e^-$ pair creation
\cite{CRV02,RV02,RV03,RVX03a,RVX03b,RVX03c}. The goal has been to clarify the
physical nature of the process of extracting the mass energy of a black
hole by the creation of $e^+e^-$ matter pairs \cite{CR71} and to analyze the electromagnetic
radiation emission process during the transient dynamical phases of the
gravitational collapse leading to the final formation of the black hole. 

In this letter we conclude this analysis by making precise predictions for the
spectra, the energy fluxes and characteristic time-scales of the radiation
for short-bursts. 
If the precise luminosity variation and spectral hardening of the radiation we
have predicted will be confirmed by observations of short-bursts, these systems
will play a major role as standard candles in cosmology.

These considerations will also be relevant for the analysis of the long-bursts when the baryonic matter contribution will be taken into account.

\section{The model}

The idea that the origin of GRBs is related to the energy extractable from a black hole \cite{CR71} by process of vacuum polarization \cite{S31,HE35,S51} and the creation of $e^+e^-$ plasma was advanced in 1974 by Damour and Ruffini \cite{DR75}. The basic considerations on the dynamics of the $e^+e^-$ plasma in the context of GRBs were outlined in 1978 by Cavallo and Rees \cite{CR78}, without addressing the issue of the origin of this plasma. In 1998 \cite{PRX98} these concepts were further evolved by the identification of the region around an already formed black hole in which such $e^+e^-$ plasma can be created and the concept of ``dyadosphere'' was introduced.

In this letter for the first time we present progress in describing the expected radiation from the dynamical formation of the dyadosphere in the process of gravitational collapse.

The dynamics of the collapse of an electrically-charged stellar core, separating
itself from an oppositely charged remnant in an initially neutral star, was
first modelled by an exact solution of the Einstein-Maxwell equations
corresponding to a shell of charged matter in Ref.~\cite{CRV02}. The
fundamental dynamical equations and their analytic solutions were obtained, 
revealing the amplification of the electromagnetic field strength
during the process of collapse and the asymptotic approach to the final static
configuration. The results, which properly account for general
relativistic effects, are summarized in Fig.~1 and Fig.~2 of
Ref.~\cite{CRV02}.

A first step toward the understanding of the process of extracting energy
from a black hole was obtained in Ref.~\cite{RV02}, where it
was shown how the extractable electromagnetic energy is not stored behind the
horizon but is actually distributed all around the black hole. Such a stored
energy is in principle extractable, very efficiently, on time-scales
$\sim\hbar/m_{e}c^{2}$, by a vacuum polarization process \emph{\`{a} l\`{a}}
Sauter-Heisenberg-Euler-Schwinger \cite{S31,HE35,S51}. Such a process occurs
if the electromagnetic field becomes larger than the critical field strength
$\mathcal{E}_{\mathrm{c}}$ for $e^+e^-$ pair creation. In
Ref.~\cite{RV02} we followed the approach of Damour and Ruffini \cite{DR75} in
order to evaluate the energy density and the temperature of the created
$e^+e^-$-photon plasma. As a byproduct, a formula for the
irreducible mass of a black hole was also derived solely in terms of the
gravitational, kinetic and rest mass energies of the collapsing core. This
surprising result allowed us in Ref.~\cite{RV03} to obtain a deeper
understanding of the maximum limit for the extractable energy during the process of
gravitational collapse, namely 50\% of the initial energy of
the star: the well known result of a 50\% maximum efficiency for energy extraction in
the case of a Reissner-Nordstr\"{o}m black hole \cite{CR71} then becomes a particular
case of a process of much more general validity.

The crucial issue
of the survival of the electric charge of the collapsing core in the presence of a
copious process of $e^+e^-$ pair creation was addressed in
Refs.~\cite{RVX03a,RVX03b}. By using theoretical techniques borrowed from
plasma physics and statistical mechanics
\cite{GKM87,KESCM91,KESCM92,CEKMS93,KME98,SBR...98,BMP...99} based on a
generalized Vlasov equation, it was possible to show that while the core keeps
collapsing, the created $e^+e^-$ pairs are entangled in the
overcritical electric field. The electric field itself, due to the back
reaction of the created $e^+e^-$ pairs, undergoes damped
oscillations in sign finally settling down to the critical value
$\mathcal{E}_{\mathrm{c}}$. The pairs fully thermalize to an
$e^+e^-$-photon plasma on time-scales typically of the order of
$10^{2}$--$10^{4}\hbar/m_{e}c^{2}$. During this characteristic damping time,
which we recall is much larger than the pair creation time-scale
$\hbar/m_{e}c^{2}$, the core moves
inwards, collapsing with a speed $0.2$--$0.8c$,
further amplifying the electric field strength at its surface and
enhancing the pair creation process.

Turning now to the dynamical evolution of such an $e^+e^-$ plasma we recall that, after some original attempt to consider a steady state emission \cite{p86,p90}, the crucial progress was represented by the understanding that during the optically thick phase such a plasma expands as a thin shell. There exists a fundamental relation between the width of the expanding shell and the Lorentz gamma factor. The shell expands, but the Lorentz contraction is such that its width in laboratory frame appears to be constant. Such a result was found in \cite{PSN93} on the basis of a numerical approach, further analyzed in Bisnovatyi-Kogan and Murzina \cite{BM95} on the basis of an analytic approach. Attention to the role of the rate equations governing the $e^+e^-$ annihilation were given in \cite{GW98}, where approximations to the full equation were introduced. These results were improved in two important respects in 1999 and 2000 \cite{RSWX99,RSWX00}: the initial conditions were made more accurate by the considerations of the dyadosphere as well as the dynamics of the shell was improved by the self-consistent solution of the hydrodynamical equation and the rate equation for the $e^+e^-$ plasma following both an analytic and numerical approach. 

We are now ready to report in this letter the result of using the approach in \cite{RSWX99,RSWX00} in this general framework describing the dynamical formation of the dyadosphere.

The first attempt to analyze the
expansion of the newly generated and thermalized $e^+e^-$-photon
plasma was made in Ref.~\cite{RVX03c}. The initial dynamical phases of the
expansion were analyzed, using the general relativistic equations of
Ref.~\cite{CRV02} for the gravitational collapse of the core. A {\itshape separatrix} was found in the motion of the plasma
at a critical radius $\bar{R}$: the plasma created at radii larger than
$\bar{R}$ expands to infinity, while the one created at radii smaller than
$\bar{R}$ is trapped by the gravitational field of the collapsing core and
implodes towards the black hole. The value of $\bar{R}$ was found in
Ref.~\cite{RVX03c} to be 
$\bar{R}=2GM/c^{2}[1+\left(  1-3Q^{2}/4GM^{2}\right)^{1/2}]$, 
where $M$ and $Q$ are the mass and the charge of the core, respectively.

In this letter we pursue further the evolution of such a
system, describing the dynamical phase of the expansion of the pulse of the
optically thick plasma all the way to the point where the transparency
condition is reached. Some pioneering work in this respect were presented in Goodman in 1986 \cite{G86}. In this process the pulse reaches ultrarelativistic
regimes with Lorentz factor $\gamma\sim10^{2}$--$10^{4}$. The spectra, the
luminosities and the time-sequences of the electromagnetic signals captured
by a far-away observer are analyzed here in detail for the first time. 
The relevance of these theoretical results for short-bursts is then discussed.

\section{The expansion of the $e^{+}e^{-}\gamma$ plasma as a discrete set of
elementary slabs}

We discretize the gravitational collapse of a spherically symmetric
core of mass $M$ and charge $Q$
by considering a set of events along the world line of a point of fixed
angular position on the collapsing core surface. Between each of
these events we consider a spherical shell slab of plasma of constant coordinate
thickness $\Delta r$ so that:

\begin{enumerate}
\item $\Delta r$ is assumed to be a constant which is 
small with respect to the core radius;

\item $\Delta r$ is assumed to be large with respect to the mean free path of
the particles so that the statistical description of the $e^{+}e^{-}\gamma$
plasma can be used;

\item  There is no overlap among the slabs and their union describes the
entirety of the process.
\end{enumerate}

We check that the final results are independent of the special value of the
chosen $\Delta r$.

 In order to describe the dynamics of the
expanding plasma pulse the energy-momentum conservation law and the rate equation
for the number of pairs in the Reissner-Nordstr\"{o}m geometry external to
the collapsing core have to be integrated:
\begin{eqnarray}
T^{\mu\nu}{}_{;\mu}  &  = &0,\label{Tab}\\
\left(  n_{e^{+}e^{-}}u^{\mu}\right)  _{;\mu}  &  = & \overline{\sigma v}\left[
n_{e^{+}e^{-}}^{2}\left(  T\right)  -n_{e^{+}e^{-}}^{2}\right]  , \label{na}%
\end{eqnarray}
where $T^{\mu\nu}=\left(  \epsilon+p\right)  u^{\mu}u^{\nu}+pg^{\mu\nu}$ is
the energy-momentum tensor of the plasma with proper energy density
$\epsilon$ and proper pressure $p$, $u^{a}$ is the fluid 4-velocity,
$n_{e^{+}e^{-}}$ is the pair number density, $n_{e^{+}e^{-}}\left(  T\right)
$ is the equilibrium pair number density at the temperature $T$ of the plasma
and $\overline{\sigma v}$ is the mean of the product of the $e^{+}e^{-}$
annihilation cross-section and the thermal velocity of the pairs. We use
Eqs.~(\ref{Tab}) and (\ref{na}) to study the expansion of each slab, following
closely the treatment developed in Refs~\cite{RSWX99,RSWX00} where it was
shown how a homogeneous slab of plasma expands as a pair-electromagnetic
pulse (PEM pulse) of constant thickness in the laboratory frame. Two regimes
can be identified in the expansion of the slabs:

\begin{figure}[th]
\begin{center}
\includegraphics[width=9cm]{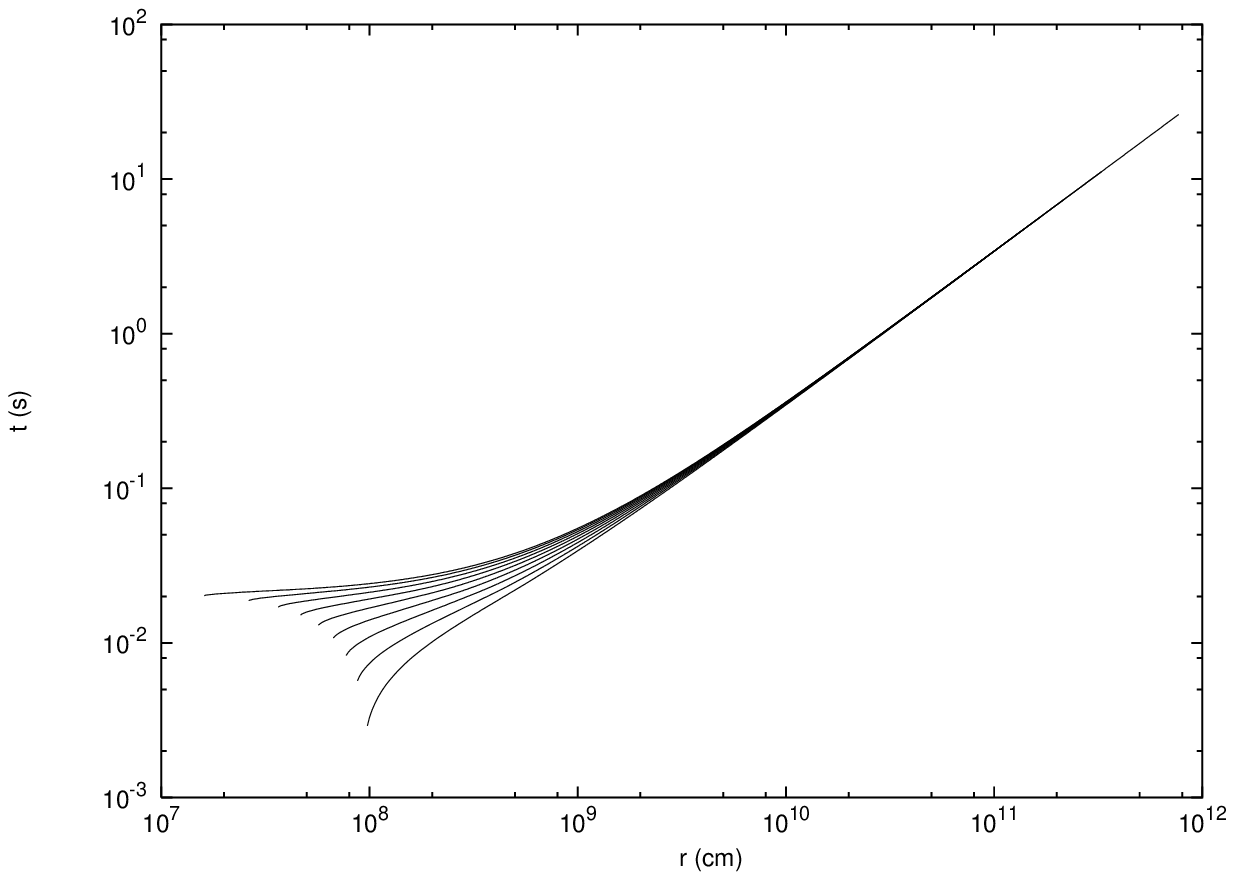} \includegraphics
[width=9cm]{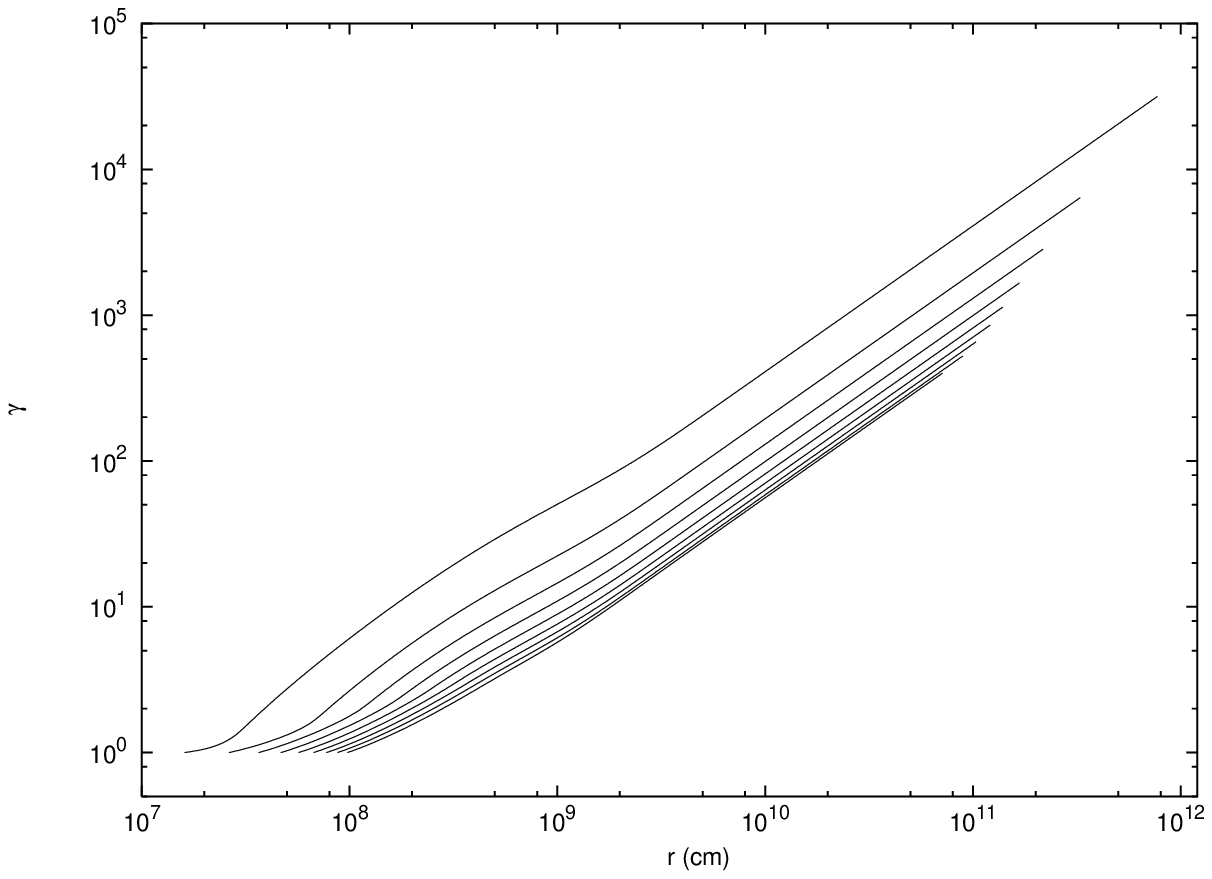}
\end{center}
\caption{Expansion of the plasma created around an overcritical collapsing
stellar core with $M=10M_{\odot}$ and $Q=0.1\sqrt{G}M$. Upper diagram: world
lines of the plasma. Lower diagram: Lorentz $\gamma$ factor as a function of
the radial coordinate $r$.}%
\label{ff1}%
\end{figure}

\begin{enumerate}
\item  In the initial phase of expansion the plasma experiences the strong
gravitational field of the core and a fully general relativistic description of
its motion is needed. The plasma is sufficiently hot in this first phase that
the $e^{+}e^{-}$ pairs and the photons remain at thermal equilibrium in it. As
shown in Ref.~\cite{RVX03c}, under these circumstances, the right hand side of
Eqs.~(\ref{na}) is effectively $0$ and Eqs.~(\ref{Tab}) and (\ref{na}) 
are equivalent to:
\begin{eqnarray}
\left(  \tfrac{dr}{cdt}\right)  ^{2}=\alpha^{4}\left[  1-\left(
\tfrac{n_{e^{+}e^{-}}}{n_{e^{+}e^{-}0}}\right)  ^{2}\left(  \tfrac{\alpha_{0}%
}{\alpha}\right)  ^{2}\left(  \tfrac{r}{r_{0}}\right)  ^{4}\right]  ,\label{Eq2}\\
\left(  \tfrac{r}{r_{0}}\right)  ^{2}=\left(  \tfrac{\epsilon+p}{\epsilon_{0}%
}\right)  \left(  \tfrac{n_{e^{+}e^{-}0}}{n_{e^{+}e^{-}}}\right)  ^{2}\left(
\tfrac{\alpha}{\alpha_{0}}\right)  ^{2}-\tfrac{p}{\epsilon_{0}}\left(
\tfrac{r}{r_{0}}\right)  ^{4},\nonumber
\end{eqnarray}
where $r$ is the radial coordinate of a slab of plasma, $\alpha=\left(
1-2MG/c^{2}r\right.$ $\left.+Q^{2}G/c^{4}r^{2}\right)^{1/2}$ is
the gravitational redshift factor and the subscript 
``$\scriptstyle{0}$" refers to quantities evaluated at the initial time.

\item  At asymptotically late times the temperature of the plasma drops below
an equivalent energy of $0.5$ MeV 
and the $e^{+}e^{-}$ pairs and the photons can no longer be
considered to be
in equilibrium: the full rate equation for pair
annihilation needs to be used. However, the plasma is so far from the central
core that gravitational effects can be neglected. In this new regime, as
shown in Ref.~\cite{RSWX99}, Eqs.~(\ref{Tab}) and (\ref{na}) reduce to:
\begin{eqnarray}
\tfrac{\epsilon_{0}}{\epsilon}  &  = & \left(  \tfrac{\gamma\mathcal{V}}%
{\gamma_{0}\mathcal{V}_{0}}\right)  ^{\Gamma},\nonumber\\
\tfrac{\gamma}{\gamma_{0}}  &  = & \sqrt{\tfrac{\epsilon_{0}\mathcal{V}_{0}%
}{\epsilon\mathcal{V}}},\label{Eq3}\\
\tfrac{\partial}{\partial t}N_{e^{+}e^{-}}  &  = & -N_{e^{+}e^{-}}\tfrac
{1}{\mathcal{V}}\tfrac{\partial\mathcal{V}}{\partial t}+\overline{\sigma
v}\tfrac{1}{\gamma^{2}}\left[  N_{e^{+}e^{-}}^{2}\left(  T\right)
-N_{e^{+}e^{-}}^{2}\right]  ,\nonumber
\end{eqnarray}
where $\Gamma=1+p/\epsilon$, $\mathcal{V}$ is the volume of a single
slab as measured in the laboratory frame by an observer at rest with the black hole, $N_{e^{+}e^{-}}=\gamma n_{e^{+}e^{-}}$ is the pair number density as
measured in the laboratory frame by an observer at rest with the black hole,
and $N_{e^{+}e^{-}}\left(  T\right)  $ is the equilibrium laboratory pair
number density.
\end{enumerate}

\section{The reaching of transparency and the signature of the outgoing
gamma ray signal}

Eqs.~(\ref{Eq2}) and (\ref{Eq3}) must be separately integrated 
and the solutions matched at
the transition between the two regimes. The integration stops when each slab
of plasma reaches the optical transparency condition given by
\begin{equation}
\int_0^{\Delta r}\sigma_{T}n_{e^{+}e^{-}}dr\sim1 \,,
\end{equation}
where $\sigma_{T}$ is the Thomson cross-section and the integral extends over
the radial thickness $\Delta r$ of the slab. The evolution of each slab occurs
without any collision or interaction with the other slabs; see the upper diagram
in Fig.~\ref{ff1}. The outer layers are colder than the inner ones and
therefore reach transparency earlier; see the lower diagram in Fig.~\ref{ff1}.
In Fig.~\ref{ff1}, Eqs.~(\ref{Eq2}) and (\ref{Eq3}) have been integrated for a
core with
\begin{equation}
M=10M_{\odot},\quad Q=0.1\sqrt{G}M;
\end{equation}
the upper diagram represents the world lines of the plasma as functions of the radius,
while the lower diagram shows the corresponding Lorentz $\gamma$ factors. The overall independence of the result of the dynamics
on the number $N$ of the slabs adopted in the discretization process or analogously on the value of $\Delta r$ has also
been checked. We have repeated the integration for $N=10$, $N=100$ reaching 
the same result to extremely good accuracy. The results in Fig.~\ref{ff1}
correspond to the case $N=10$.

\begin{figure}[th]
\begin{center}
\includegraphics[width=9cm]{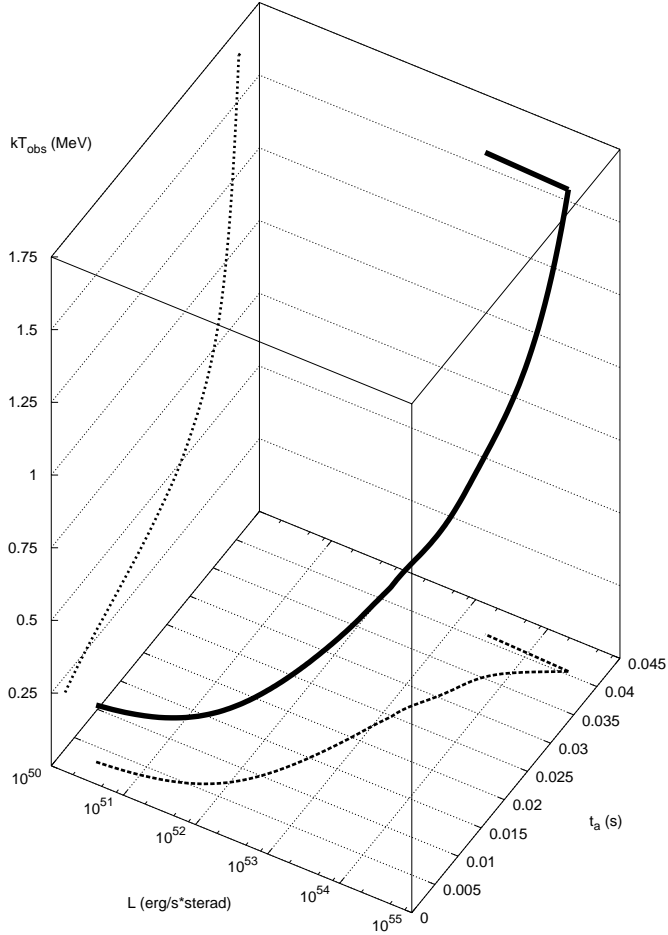}
\end{center}
\caption{Predicted observed luminosity and observed spectral hardness of the
electromagnetic signal from the gravitational collapse of a collapsing core
with $M=10M_{\odot}$, $Q=0.1\sqrt{G}M$ at $z=1$ as functions of the arrival
time $t_{a}$.}%
\label{ff4}%
\end{figure}

 We now turn to the results in
Fig.~\ref{ff4}, where we plot both the theoretically predicted luminosity $L$
and the spectral hardness of the signal reaching a far-away observer as
functions of the arrival time $t_{a}$. Since all three of these quantities
depend in an essential way on the cosmological redshift factor $z$, see
Refs.~\cite{BRX01,RBCFX03}, we have adopted a cosmological
redshift $z=1$ for this figure.

 As the plasma becomes transparent, gamma ray
photons are emitted. The energy $\hbar\omega$ of the observed photon is
$\hbar\omega=k\gamma T/\left(  1+z\right)  $, where $k$ is the Boltzmann
constant, $T$ is the temperature in the comoving frame of the pulse and
$\gamma$ is the Lorentz factor of the plasma at the transparency time. We also
recall that if the initial zero of time is chosen as the time when the first photon
is observed, then the arrival time $t_{a}$ of a photon at the detector in
spherical coordinates centered on the black hole is given by
\cite{BRX01,RBCFX03}:
\begin{equation}
t_{a}=\left(  1+z\right)  \left[  t+\tfrac{r_{0}}{c}-\tfrac{r\left(  t\right)
}{c}\cos\theta\right]
\end{equation}
where $\left(  t,r\left(  t\right)  ,\theta,\phi\right)  $ labels the
laboratory emission event along the world line of the emitting slab and
$r_{0}$ is the initial position of the slab. The projection of the plot in
Fig.~\ref{ff4} onto the $t_{a}$-$L$ plane gives the total luminosity as the
sum of the partial luminosities of the single slabs. The sudden decrease of
the intensity at the time $t=0.040466$ s corresponds to the creation of the
{\itshape separatrix} introduced in Ref.~\cite{RVX03b}. We find that the duration of the
electromagnetic signal emitted by the relativistically expanding pulse is
given in arrival time by
\begin{equation}
\Delta t_{a}\sim5\times10^{-2}\mathrm{s}\label{ta1} \,.
\end{equation}
The projection of the plot in Fig.~\ref{ff4} onto the $k T_{\mathrm{obs}%
}$, $t_{a}$ plane describes the temporal evolution of the spectral hardness.
We observe a precise soft-to-hard evolution of the spectrum of the gamma ray signal from $\sim10^{2}$ KeV monotonically increasing to $\sim1$ MeV. We
recall that $kT_{\mathrm{obs}}=k\gamma T/\left(  1+z\right)  $. 

 The
above quantities are clearly functions of the cosmological redshift $z$, of
the charge $Q$ and the mass $M$ of the collapsing core. We present in Fig.~3
the arrival time interval for $M$ ranging from $M\sim10M_{\odot}$ to
$10^{3}M_{\odot}$, keeping $Q=0.1\sqrt{G}M$. The arrival time interval is very
sensitive to the mass of the black hole:
\begin{equation}
\Delta t_{a}\sim10^{-2}-10^{-1}\mathrm{s} \,.
\label{ta2}
\end{equation}
Similarly the spectral hardness of the signal is sensitive to the ratio
$Q/\sqrt{G}M$ \cite{RFVX04}. Moreover the duration, the spectral hardness and
luminosity are all sensitive to the cosmological redshift $z$ (see
Ref.~\cite{RFVX04}).
All the above quantities can also be sensitive to a possible baryonic
contamination of the plasma due to the remnant of the progenitor star which
has undergone the process of gravitational collapse.

\begin{figure}[th]
\begin{center}
\includegraphics[width=9cm]{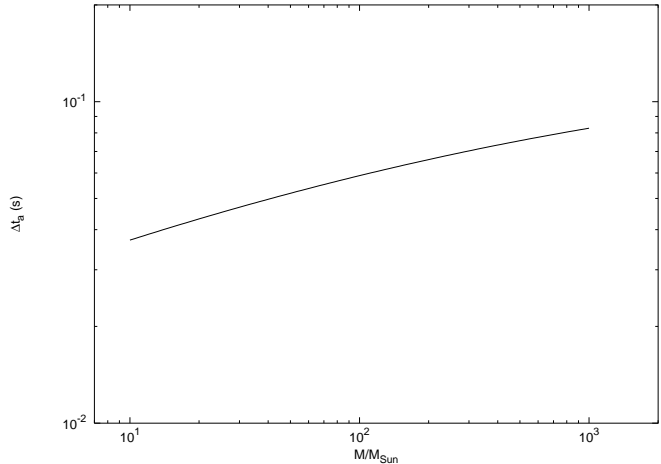}
\end{center}
\caption{Arrival time duration of the electromagnetic signal from the
gravitational collapse of a stellar core with charge $Q=0.1\sqrt{G}M$ as a
function of the mass $M$ of the core.}%
\label{ff5}%
\end{figure}

\section{Conclusions}

The above results were obtained considering $e^+e^-$ plasma without any baryonic contamination and are therefore directly relevant for short-bursts \cite{lett2}.
The characteristic spectra, time variabilities and luminosities of the
electromagnetic signals from collapsing overcritical stellar cores, here
derived from first principles, agrees very closely with the observations of
short-bursts \cite{P...99}. New space missions must be planned, with temporal resolution down
to fractions of $\mu$s and higher collecting area and spectral resolution than at present,
in order to verify the detailed agreement between our model
and the observations. It is now clear that if our theoretical
predictions will be confirmed, we would have a very powerful tool for
cosmological observations: the independent information about luminosity,
time-scale and spectrum can uniquely determine the mass, the electromagnetic
structure and the distance from the observer of the collapsing core, see e.g.\
Fig.~\ref{ff5} and Ref.~\cite{RFVX04}. In that case short-bursts may become the
best example of standard candles in cosmology \cite{R03}. The introduction we are currently analysing is the introduction of baryonic matter in the optically thick phase of the expansion of the $e^+e^-$ plasma which can affect the structure of the Proper-GRB (P-GRB) \cite{lett1} as well as the structure of the long-bursts \cite{Brasile}.

An interesting proposal was advanced in 2002 \cite{it02} that the $e^+e^-$ plasma may have a fundamental role as well in the physical process generating jets in the extragalactic radio sources. The concept of dyadosphere originally introduced in Reissner-Nordstr\"{o}m black hole in order to create the $e^+e^-$ plasma relevant for GRBs can also be generalized to the process of vacuum polarization originating in a Kerr-Newman black hole due to magneto-hydrodynamical process of energy extraction (see e.g. \cite{p01} and references therein). The concept therefore introduced in this letter becomes relevant for both the extraction of rotational and electromagnetic energy from the most general black hole \cite{CR71}.

After the submission of this letter we have become aware that Ghirlanda et al. \cite{g03} have given evidence for the existence of an exponential cut off at high energies in the spectra of short bursts. We are currently comparing and contrasting these observational results with the predicted cut off in Fig.~\ref{ff4} which results from the existence of the separatrix introduced in \cite{RVX03a}. The observational confirmation of the results presented in Fig.~\ref{ff4} would lead for the first time to the identification of a process of gravitational collapse and its general relativistic self-closure as seen from an asymptotic observer.

\end{document}